\documentclass{WileyMSP-template}

\usepackage{graphicx}
\usepackage[numbers,sort&compress,super,square]{natbib}
\usepackage{CJK}
\usepackage{amsmath,amssymb}
\usepackage[normalem]{ulem}
\usepackage{siunitx}
\usepackage[caption=false]{subfig}
\graphicspath{ {./figures/} }
\usepackage{ifpdf}\ifpdf 
    \usepackage[pdftitle={Microlayer in nucleate boiling as a Landau-Levich film},linktocpage,pdftex,bookmarksopen,bookmarksopenlevel=1,hyperfigures]{hyperref}
\else
    
    \usepackage{hyperref}
\fi
\begin{document}
\pagestyle{fancy}
\rhead{\includegraphics[width=2.5cm]{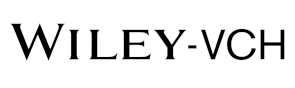}}
\title{\noindent Microlayer in nucleate boiling seen as Landau-Levich film with dewetting and evaporation\\}
\maketitle
\author{Cassiano Tecchio*}
\author{Xiaolong Zhang}
\author{Benjamin Cariteau}
\author{Gilbert Zalczer}
\author{Pere Roca i Cabarrocas}\\
\author{Pavel Bulkin}
\author{J\'er\^ome Charliac}
\author{Simon Vassant}
\author{Vadim S. Nikolayev*}\\
\begin{CJK*}{UTF8}{gbsn}
\begin{affiliations}
Dr. C. Tecchio, Dr. B. Cariteau\\
STMF, Universit\'e Paris-Saclay, CEA, 91191 Gif-sur-Yvette Cedex, France\\
cassiano.tecchio@cea.fr\\
Dr. X. Zhang (张晓龙), Dr. G. Zalczer, Dr. S. Vassant, Dr. V. S. Nikolayev\\
SPEC, CEA, CNRS, Universit\'e Paris-Saclay, 91191 Gif-sur-Yvette Cedex, France\\
vadim.nikolayev@cea.fr\\
Prof. P. Roca i Cabarrocas, Dr. P. Bulkin, J. Charliac\\
LPICM, CNRS, Ecole Polytechnique, Institut Polytechnique de Paris, 91120 Palaiseau, France\\
\end{affiliations}
\date{\today}

\begin{abstract}
\noindent Both experimental and theoretical studies on the microscale and fast physical phenomena occurring during the growth of vapor bubbles in nucleate pool boiling are reported. The focus is on the liquid film of micrometric thickness (``microlayer'') that can form between the heater and the liquid-vapor interface of a bubble on the millisecond time scale. The microlayer strongly affects the macroscale heat transfer and is thus important to be understood. It is shown that the microlayer can be seen as the Landau-Levich film deposited by the bubble foot edge during its receding when the bubble grows. The microlayer profile measured with white-light interferometry, the temperature distribution over the heater, and the bubble shape were observed with synchronized high-speed cameras. The microlayer consists of two regions: a ridge near the contact line followed by a longer and flatter part. The ridge could not be measured because of the intrinsic limitation of interferometry, which is analyzed. The simulations show that the ridge grows over time due to collection of liquid at contact line receding, the theoretical dynamics of which agrees with the experiment. The flatter part of the microlayer is bumped and its physical origin is explained.
\end{abstract}
\end{CJK*}


\section{Introduction}
\begin{figure}[ht]
	\centering
	\includegraphics[width=7.5cm]{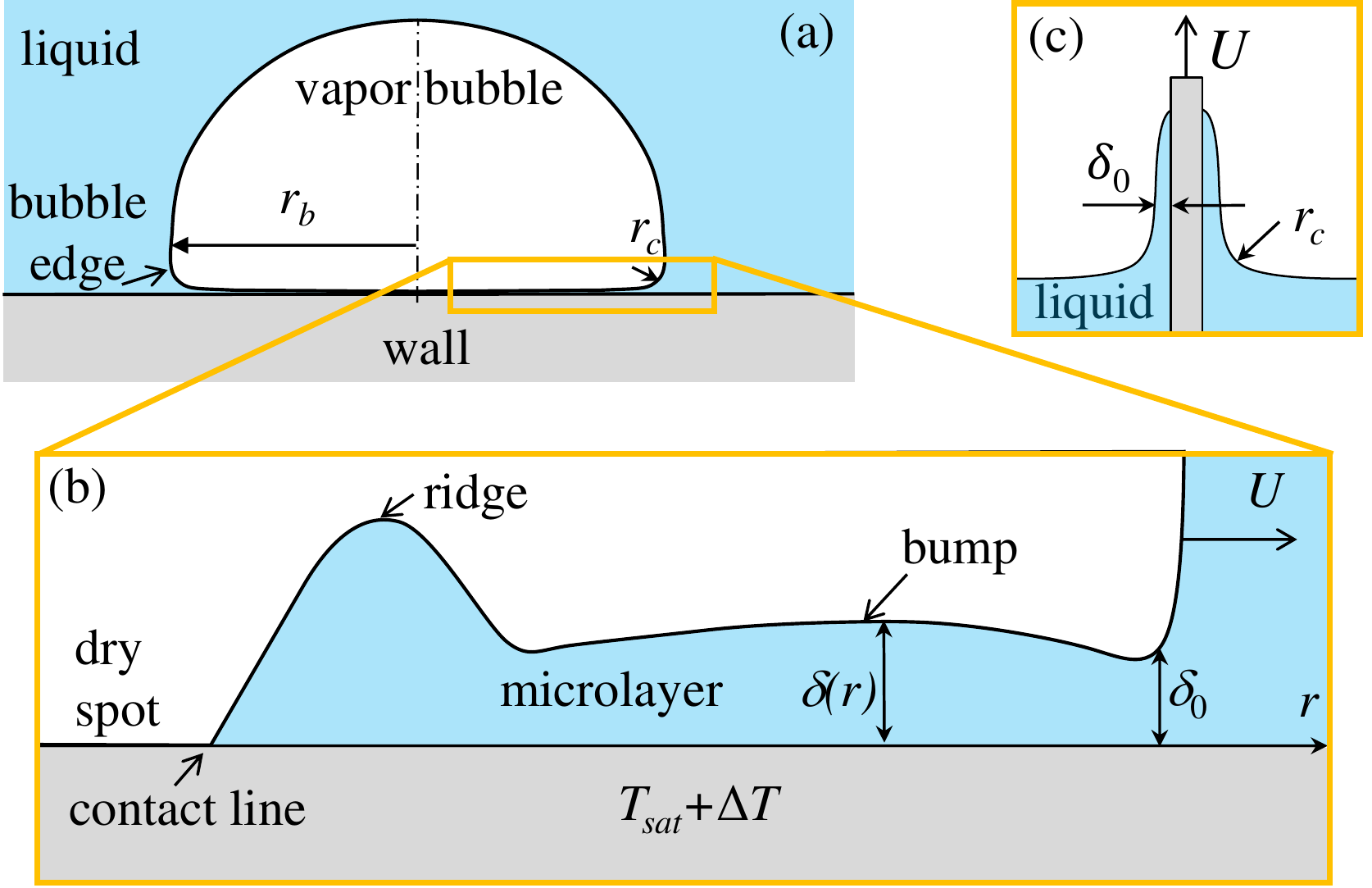}
	\caption{Schematics of the (a) general bubble shape (b) the microlayer geometry, and (c) a liquid layer deposition by dip coating.}
	\label{fig:Microlayer}	
\end{figure}

The high heat transfer rate associated to the phase change makes nucleate boiling the most suitable mode of heat transfer in a variety of industrial applications such as electronics cooling, nuclear power reactors and chemical processes. During the growth of bubbles on the heated wall, a few \si{\micro m} thick layer of liquid (known as microlayer) can be formed between the wall and the liquid-vapor interface of the bubble,\cite{Cooper69} cf. \autoref{fig:Microlayer}a,b. Its extent on the wall is up to a few \si{mm}. Heat is then transferred from the heated wall towards the liquid-vapor interface through the microlayer. This heat transfer promotes the microlayer evaporation and its contribution to the bubble growth is important in many boiling configurations.\cite{Kim09,Bongarala22} As the microlayer drying leads to the dry spot spreading, the knowledge of its dynamics is vital for the clear understanding of the boiling crisis (critical heat flux, CHF, i.e. the maximum heat flux transferrable with nucleate boiling).\cite{Zhang23b}

The most complete information about the microlayer is obtained by its direct observation. In the past it was realized with the laser interference (LI) method in the single-bubble experiments.\cite{Jawurek69,Voutsinos75,Koffman83,Gao13,Jung14,Chen17,Utaka18,Zou18,Jung18,Narayan21} The main subjects of interest were the spatial distributions of two quantities: the microlayer evaporation rate and its initial thickness $\delta_0$ (\autoref{fig:Microlayer}b). The latter quantity is defined as the microlayer thickness right after its formation at the bubble edge position $r_b$. As the microlayer forms during the bubble base expansion, one can express $\delta_0$ either as a function of time $t$ or radial distance $r=r_b(t)$ where the microlayer forms.

The pioneering LI observations\cite{Jawurek69,Voutsinos75,Koffman83} were realized in 1960's--1980's. More recently, there was a regain of interest from many research groups\cite{Gao13,Jung14,Chen17,Utaka18,Zou18,Jung18,Narayan21} related to the availability of high resolution fast cameras that provide far better accuracy. The observed microlayer shape depended mostly on the heating mode. In most works the heating was homogeneous and the observed initial microlayer thickness was a monotonous function of $r$. In some works,\cite{Chen17,Utaka18} the heating was performed by directing a jet of hot gas to the bottom of the heating wall, so the heating was localized in a spot smaller than the bubble departure size. In this case, a bump was observed in the microlayer shape. It should be noted that the bump was also observed in some cases at homogeneous heating.\cite{Jung18} The origin of this bumped profile has remained largely unexplained.

The physical origin of microlayer and its thickness were studied very early on. It was understood that the microlayer appears thanks to the rapid bubble growth and the inertial force that pushes the liquid outward as the bubble expands.\cite{Cooper69} The semi-empirical formula
\begin{equation}\label{eq_Cooper}
\delta_0\propto\sqrt{\nu t}
\end{equation}
was proposed for the initial thickness, where $\nu$ is the liquid kinematic viscosity. The dependence on density means that the initial microlayer thickness is controlled by the inertial effects (in addition to viscosity). A theoretical approach based on a similar idea was developed later.\cite{Smirnov75} However, from the fluid dynamics perspective, the microlayer is a thin viscous film with a free surface. It is well known that the flow in thin films is controlled by the surface tension and viscosity only. So should be the microlayer formation.\cite{Hansch16} The flow can only exist near the bubble edge and the contact line; the flow in between is extremely weak.\cite{JFM21} For this reason, after their formation, the thickness of the micrometric films varies (decreases) due to evaporation only.

In the same vein, \citet{Schweikert19} hypothesized that the microlayer formation is analogous to the liquid film deposited during the dip coating when a flat plate is being pulled out vertically from a liquid pool with a speed $U$ (\autoref{fig:Microlayer}c). As derived by \citet{LL42}, the film thickness in this case is
\begin{equation} \label{eq:delta0}
	\delta_0=1.34r_cCa^{2/3},
\end{equation}
where $r_c$ is the radius of curvature of liquid meniscus, and $Ca=\mu U/\sigma$ is the capillary number. Here, $\mu$ and $\sigma$ are the liquid shear viscosity and surface tension, respectively. The numerical constant is the same as in the \citet{Bretherton} description of a thin film deposited by the meniscus receding in a capillary tube. This hypothesis has been recently adopted by \citet{Bures21} to propose a criterion of microlayer formation. In this work, we apply a novel -- for this domain -- experimental technique, the spectrally-resolved white-light interferometry\cite{Glovnea03} (WLI), to measure the microlayer shape and show that the microlayer can indeed be seen as the Landau-Levich film. We also describe the growth dynamics of the dry spot that forms under the growing bubble on the heater and explain the details of microlayer structure.

\section{Maximum observable interface slope}\label{maxSlope}

The microlayer thickness is determined by interferometry. It is well known that the difference in microlayer thickness between neighboring maximum and minimum of interference\cite{Jung14}
\begin{equation}\label{eq:lambda4n}
\Delta\delta=\frac{\lambda}{4n\cos\alpha},
\end{equation}
for the incident light of a wavelength $\lambda$ and the liquid refraction index $n$; $\alpha$ is the refraction angle into the liquid, which is zero in our study as the light incidence is normal. 
The vapor-liquid interface slope $\theta$ is defined with $\tan\theta=\Delta\delta/\Delta r$, where the distance $\Delta r$ corresponds to the horizontal distance between the interfacial points where the thicknesses correspond to the neighboring maximum and minimum.  This distance is related to the distance $\Delta r_p$ measured in pixels at the interferometric image with the relation $\Delta r=R\Delta r_p$, where $R$ is the inverse spatial resolution of the optical system, i.e. the physical size of \SI{1}{px}. One cannot distinguish the maximum and minimum when $\Delta r_p<\SI{1}{px}$, which results in the limitation for the observable interface slope
\begin{equation}\label{eq:thetacr}
\theta<\theta_{cr};\quad\tan\theta_{cr}=\frac{\lambda}{4nR\cos\alpha}.
\end{equation}
This limitation that applies to any kind of interferometry, is quite restrictive in our case where $R\simeq\SI{14}{\micro m/px}$ so $\theta_{cr}\lesssim 0.4^\circ$. This should be taken into consideration while considering the results in \autoref{secbump}.

\begin{figure}
\centering
\begin{minipage}{.45\linewidth}
  \includegraphics[width=\linewidth,clip]{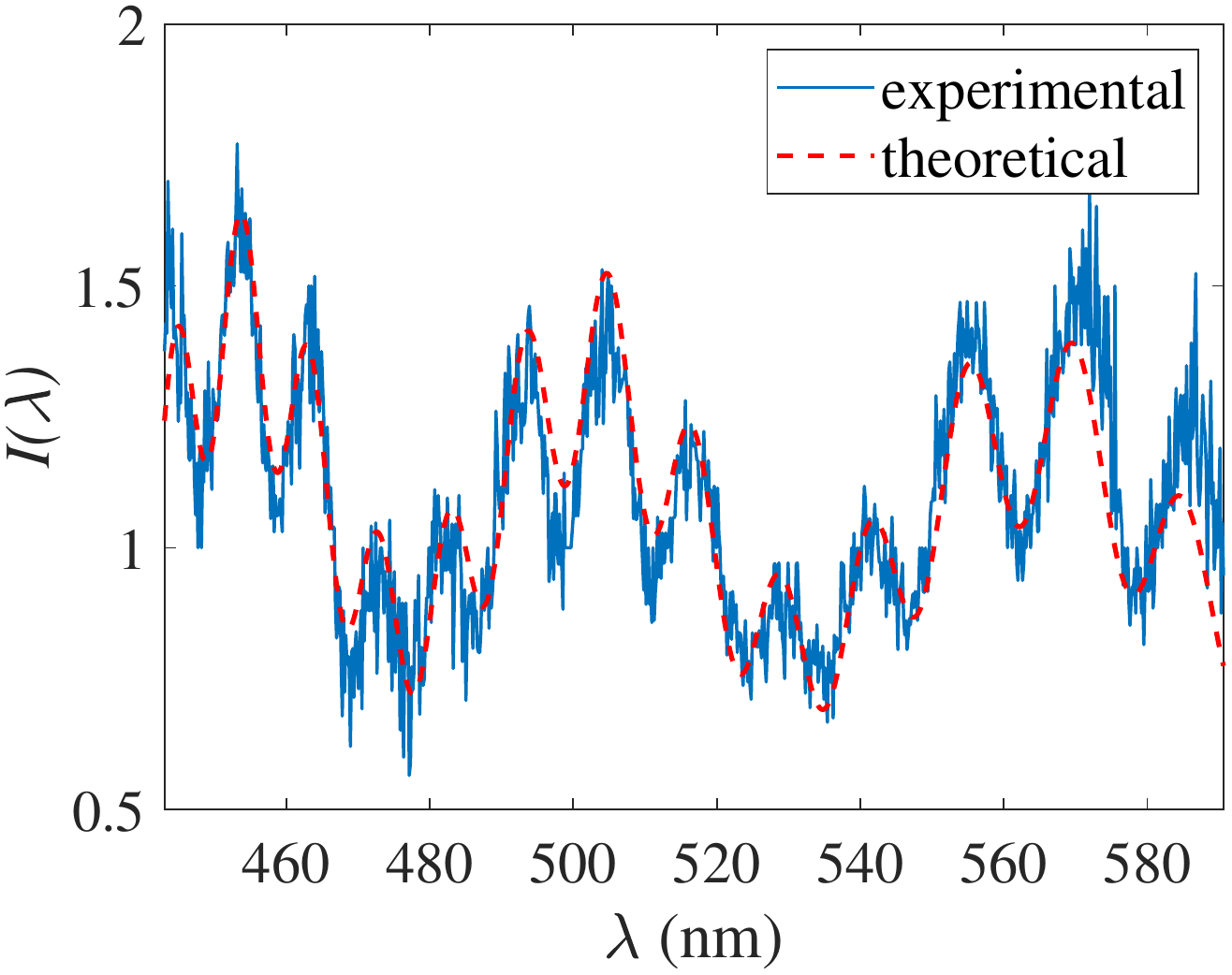}
  \caption{An example of a fit of theoretical $I_{theo}(\lambda)$ profile to the experiment  for a fixed $r$ resulting in $\delta=\SI{6.48}{\mu m}$; the ITO thickness is \SI{944}{nm}.}
  \label{fig:minimizeBoiling}
\end{minipage}
\hspace{.05\linewidth}
\begin{minipage}{.45\linewidth}
  \includegraphics[width=\linewidth,clip]{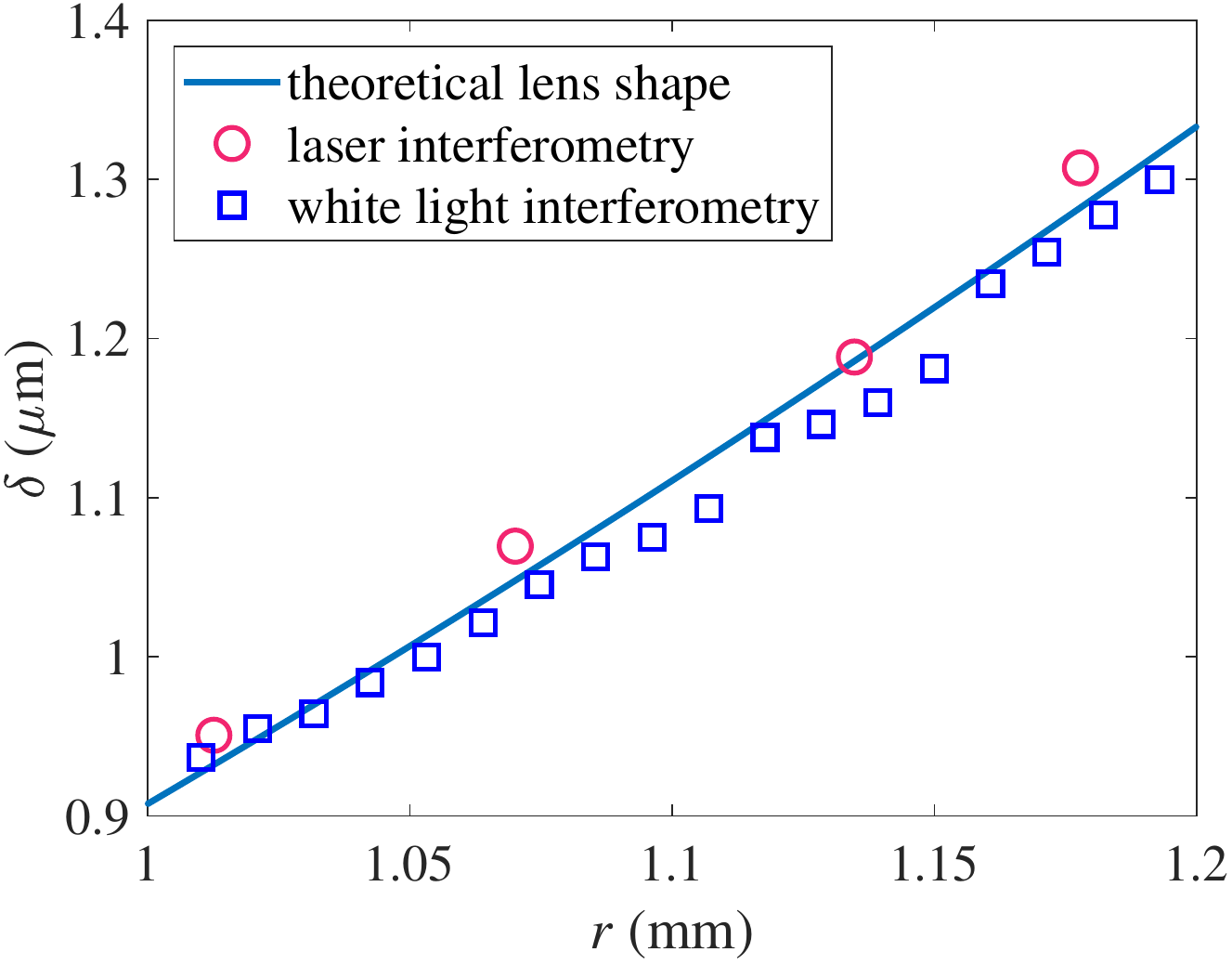}
  \caption{Comparison of LI and WLI for the validation case of air film between the ITO surface and a spherical lens posed on it. The theoretical film shape is calculated from the known lens curvature.}
  \label{fig:resolutionA}
\end{minipage}
\end{figure}

We apply here WLI because it is in many respects superior to LI. In all the previous works on microlayer, the thickness was determined by LI from the absolute number of interference maxima and minima. This requires knowing the absolute fringe number. However, typically, the slope is large near the contact line.\cite{Guion18,Urbano18,Giustini20a,Bures21a} The first fringes can thus be lost because of the maximum observable slope limitation (\autoref{eq:thetacr}), which induces a systematic error on the thickness determination for all measured points. We use here the spectrally resolved WLI, where such an error is excluded. The thickness at a given point $r$ is determined by minimizing the deviation between the experimental intensity $I_{exp}(r, \lambda)$ (cf. \autoref{ExpSec}) and the theoretical intensity\cite{Tecchio22} $I_{theo}(r, \lambda)$, see \autoref{fig:minimizeBoiling}. Such a data reduction procedure requires a careful calibration and validation. The validation was performed by measuring the profile of the air film formed between a flat surface and a lens of a known (from manufacturer) curvature (\autoref{fig:resolutionA}). The decisive advantage of WLI with respect to LI is a much larger amount of information about the profile. Indeed, WLI can be seen as LI performed simultaneously with numerous lasers of different wavelengths corresponding to the number of pixels along the $\lambda$ axis.

\section{Bumped shape of the initial microlayer thickness}\label{secbump}

\begin{figure}
\centering
\begin{minipage}{.45\linewidth}
  \includegraphics[width=0.8\linewidth,clip]{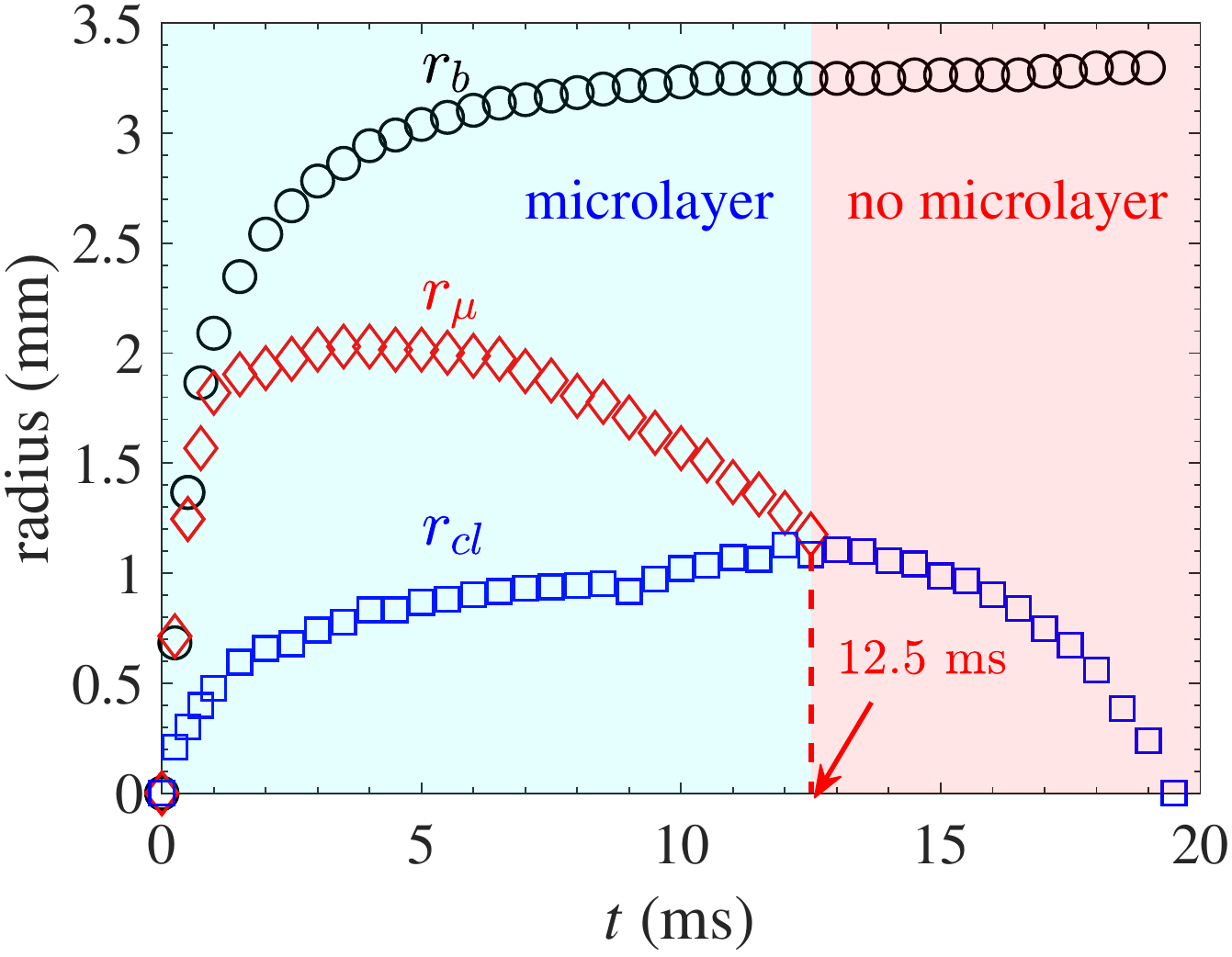}
	\caption{Experimentally measured time evolution of the bubble diameter $r_b$, microlayer radius $r_\mu$ and dry spot (contact line) radius $r_{cl}$. Their geometrical definition is shown in \autoref{fig:Setup} below.} \label{fig:radii}
\end{minipage}
\hspace{.05\linewidth}
\begin{minipage}{.45\linewidth}
	\includegraphics[width=0.8\linewidth]{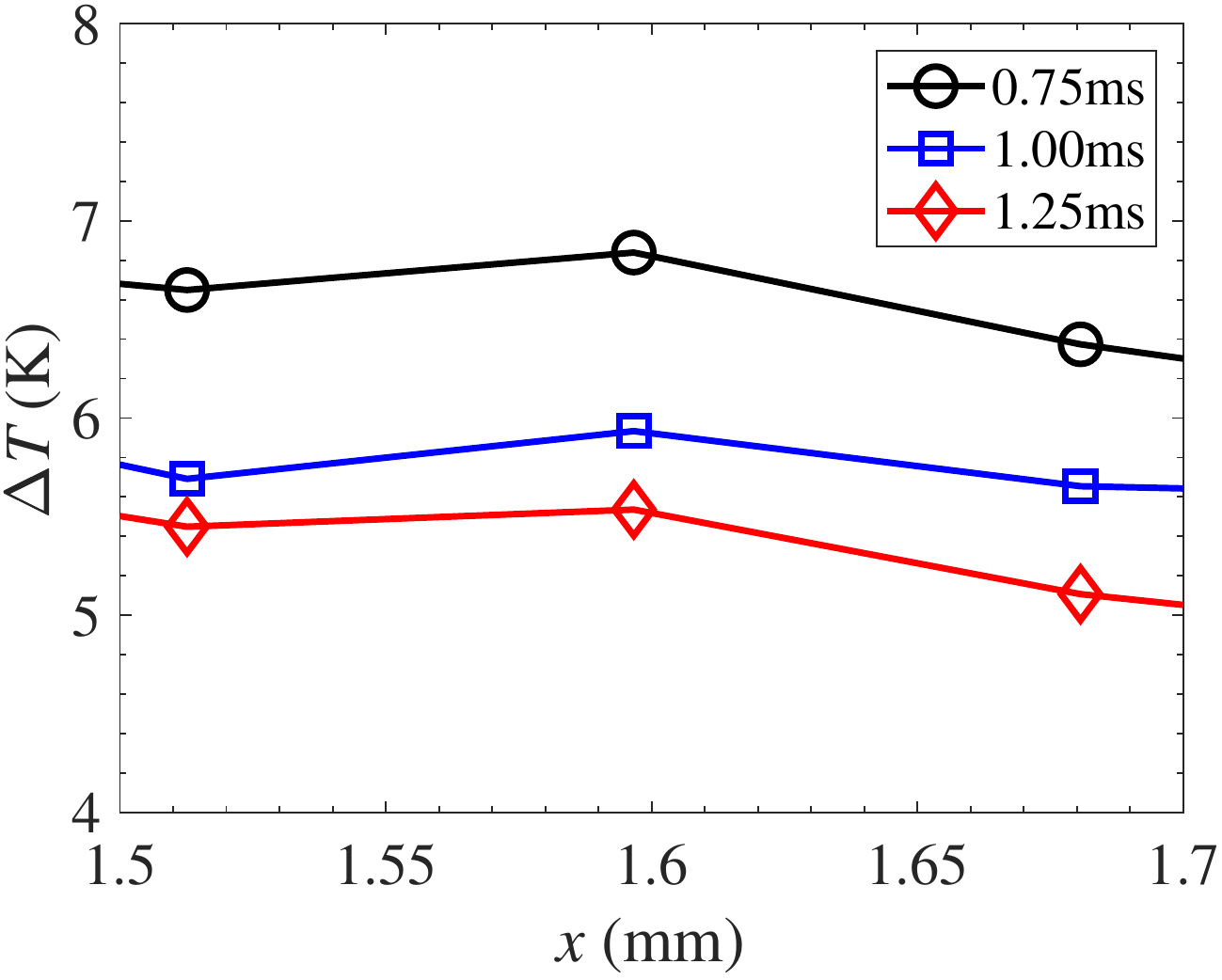}
	\caption{The wall superheating variation in the central part of microlayer where WLI measurements were performed.}
	\label{fig:DTm}	
\end{minipage}
\end{figure}

To complete an analogy between the dip coating and the microlayer formation, one needs to define a speed $U$ and a radius of curvature $r_c$. As the microlayer formation can be seen as a film deposited by the receding bubble base edge, $U=d r_b/d t$, which is the bubble edge velocity relative to the wall, see the analogy between \autoref{fig:Microlayer}a and \autoref{fig:Microlayer}c. The meniscus radius of curvature $r_c$ (\autoref{fig:Microlayer}a) can be associated to that of the bubble edge (\autoref{fig:Microlayer}c). Note that $r_c$ is different from the bubble maximum radius $r_b$. The bubble foot is flattened by the inertial forces acting on the bubble downwards thanks to the rapid bubble growth. Therefore, the portion of bubble interface between the foot and the dome is strongly curved, see \autoref{fig:Microlayer}a; $\beta=r_c/r_b$ is thus expected to be much smaller than unity. We assume it to be constant during the microlayer formation. To apply \autoref{eq:delta0} to the initial microlayer thickness, the time evolution of $r_b=r_b(t)$ is required. It is obtained experimentally (\autoref{fig:radii}) by using the sideview shadowgraphy (\autoref{ExpSec}).

One needs now to obtain $\delta_0$ from the experiment to compare with the above theory. Because of the limitation discussed in \autoref{maxSlope}, $\delta$ could be measured experimentally only in a short range of $r$. The bubble growth was however so quick that the interference fringes were not clearly distinguishable until the growth time $t_\mu=\SI{1.25}{ms}$. These perturbations are probably due to the hydrodynamic flows in microlayer in the vicinity of the bubble edge: it should displace far enough from the observation range to detect clearly the fringes. To find $\delta_0(r)$ (i.e the thickness before evaporation starts) from $\delta(r, t_\mu)$, one needs to solve thus a kind of the inverse problem.

As mentioned above, in the middle of microlayer, the fluid flow is absent so $\delta$ varies because of evaporation only. Under the conventional assumptions\cite{SWEP22} (\autoref{ExpSec}) for the description of flat micrometric liquid layers, the energy balance at the interface leads to
\begin{equation}\label{eq:energybalance}
	\mathcal{L}\rho_l\frac{\partial\delta}{\partial t} = -k\frac{\Delta T}{\delta},
\end{equation}
where $\mathcal{L}$, $\rho_l$, $k$ and stand for the latent heat, density and thermal conductivity of the liquid. $\Delta T=T_w-T_{sat}$ is the wall superheating with $T_w$ and $T_{sat}$ representing the wall and saturation temperatures, respectively; the latter is taken for the atmospheric pressure. To simplify the solution, one can use the weakness of both spatial and temporal $\Delta T$ variation along the microlayer (\autoref{fig:DTm}); we thus use a constant value $\Delta T\simeq\SI{6}{K}$. By integrating \autoref{eq:energybalance}, one obtains
\begin{equation}\label{eq:initialthickness}
	\delta_{0}(r)=\sqrt{\delta(r,t_\mu)^2+\frac{2k\Delta T}{\mathcal{L} \rho_l}t_e(r)},
\end{equation}
where $t_e$ is the time of evaporation of microlayer at its point $r$. During the time $t_e$, the bubble edge moves from the position $r$ to $r_b(t_\mu)$, so $t_e\sim\SI{0.5}{ms}$ can be found from the $r_b(t)$ curve (\autoref{fig:radii}) for each $r$.

Figure~\ref{fig:theoretical} depicts a comparison of $\delta_0$ given by the theory based on \autoref{eq:delta0} with the experimental value given by \autoref{eq:initialthickness}. The unknown \emph{a priori} value of  $\beta=0.084$ is chosen to fit the maximum experimental thickness. It is remarkable that the position of maximum thickness $r\approx\SI{1.6}{mm}$ is well reproduced by such a simple theory, which shows its validity.

One can now give a simple explanation of the bump. During the microlayer formation, $U^{2/3}$ decreases in time (in other words, the growth of $r_b$ slows down, which is a common feature of the bubble growth, cf. \autoref{fig:radii}). At the same time, $r_c\propto r_b$ increases. As $\delta_0$ is defined by their product according to \autoref{eq:delta0}, an extremum (i.e. the bump) should appear at some time. 

\section{Microlayer time evolution and contact line dynamics}

In the present experiment, the microlayer forms very quickly because of the rapid initial bubble expansion caused by strongly localized heating.
The dynamics of microlayer area can be described by considering the evolution of the microlayer radius $r_\mu$ (i.e the radius of a thin layer showing the interference fringes, cf. \autoref{fig:Setup} below for example images). It is presented in \autoref{fig:radii} together with the evolution of both $r_b$ and the contact line radius $r_{cl}$. In the beginning of bubble growth, the microlayer evolution is so quick that it is completely formed after the first $\SI{1.25}{ms}$. The contact line receding is fast but slower than the $r_\mu$ evolution. Until $\SI{5}{ms}$, the bubble edge continues to slowly recede (i.e. moves toward the liquid bulk) so $r_\mu$ keeps slowly growing. For $\SI{5}{ms}<t<\SI{12.5}{ms}$, the bubble edge changes its direction of motion and advances over the microlayer until the latter is completely depleted. Note that its depletion occurs not because of  its evaporation but because of its length reduction. This feature is commonly observed in contemporary microlayer studies.\cite{Jung14}

\begin{figure}[thb]
	\centering
\begin{minipage}{.45\linewidth}
	\includegraphics[width=0.8\linewidth]{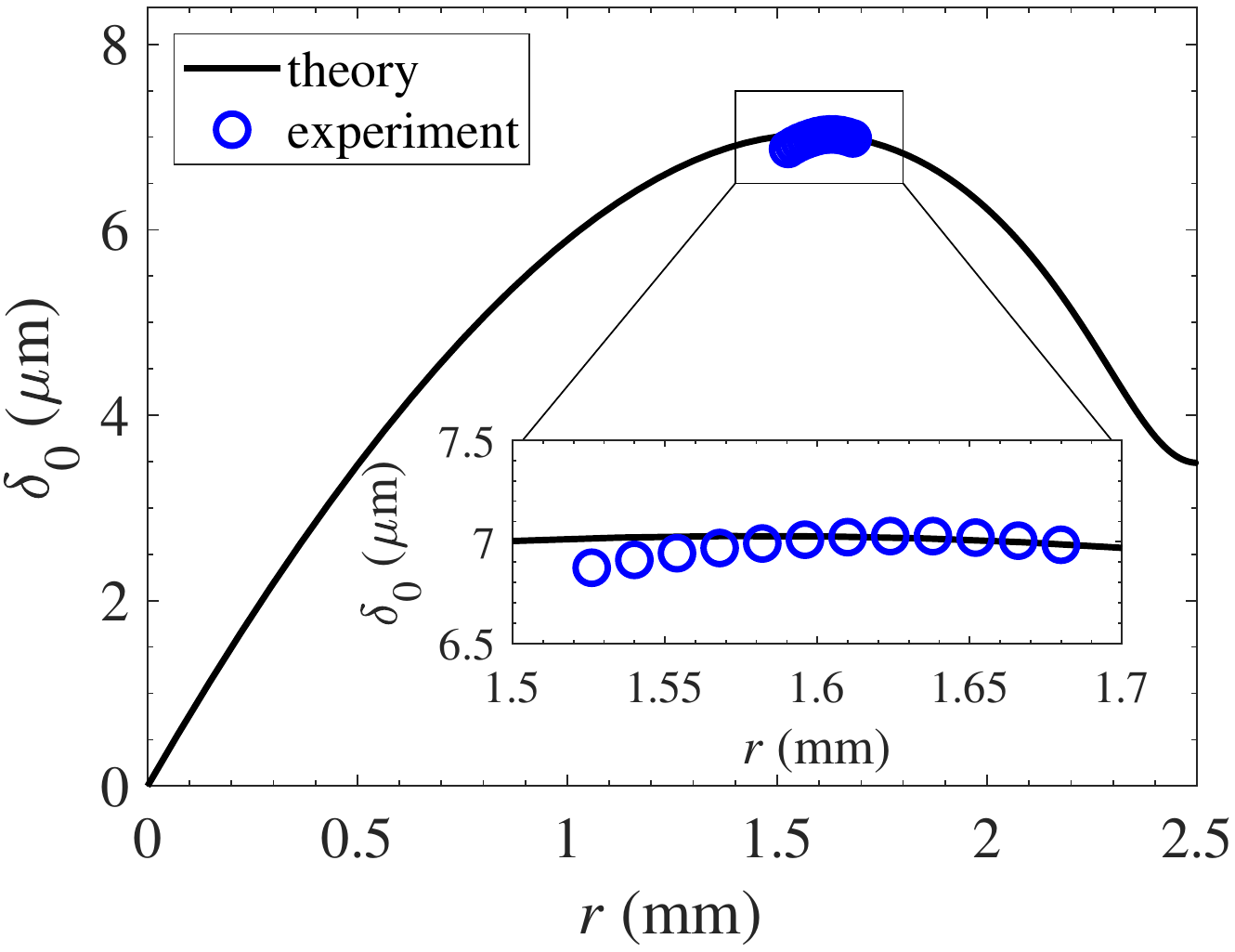}
	\caption{Initial microlayer profile. The experimental WLI data and theory based on \autoref{eq:delta0} are compared.}
	\label{fig:theoretical}	
\end{minipage}
\hspace{.05\linewidth}
\begin{minipage}{.45\linewidth}
	\includegraphics[width=8.5cm]{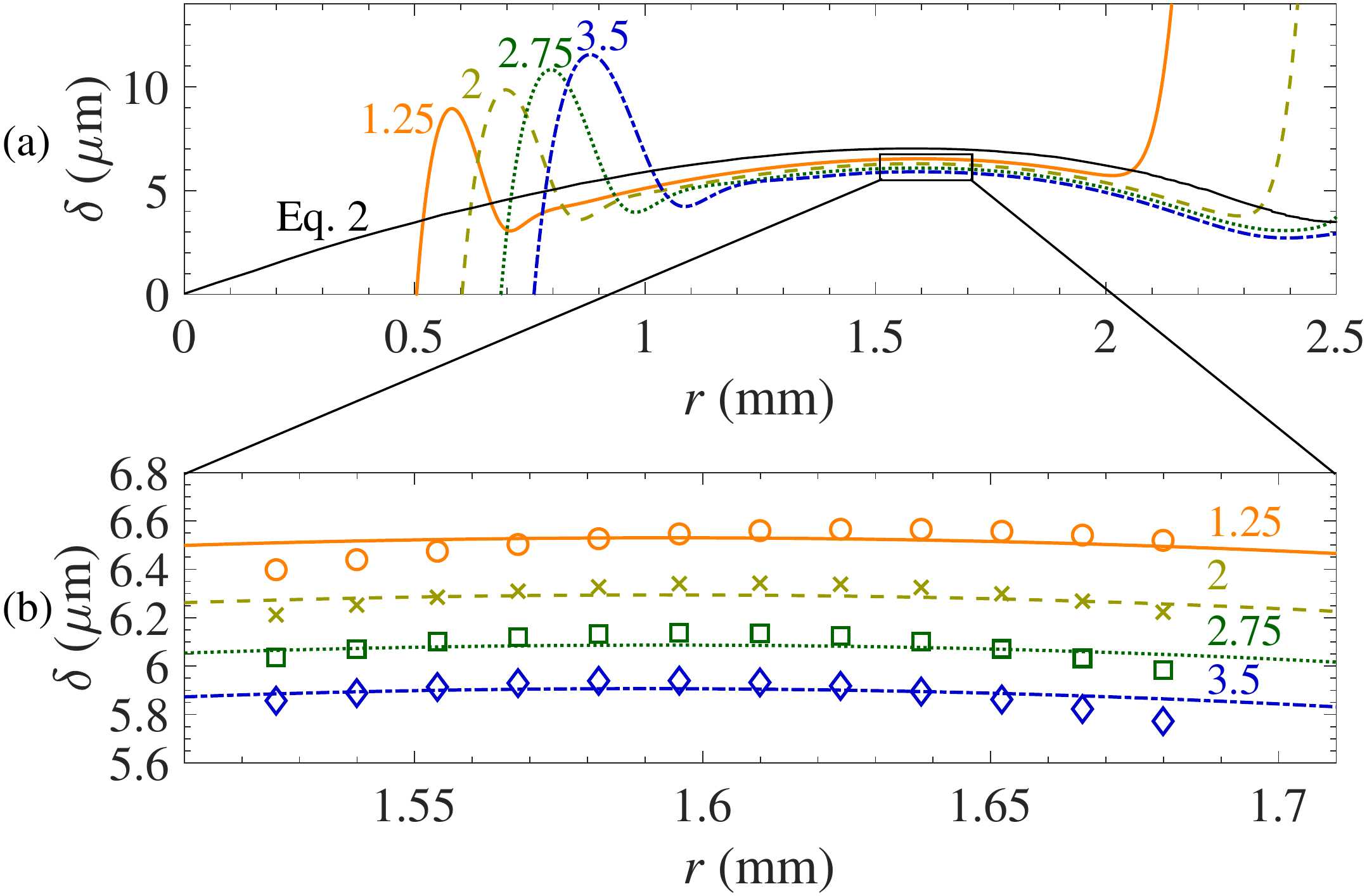}
	\caption{(a) Temporal evolution of the numerical microlayer profile. See the Supporting information (SI) for the corresponding video. The initial microlayer thickness  given by \autoref{eq:delta0} is also shown. (b) Comparison between numerical simulation (lines) and experimental WLI data (characters). The corresponding time is labeled in ms.}
	\label{fig:simul}	
\end{minipage}
\end{figure}

In order to interpret and understand the experimental observations, a numerical simulation of the microlayer profile is performed using a two-dimensional generalized lubrication approximation,\cite{JFM22,EPL23} cf. \autoref{ExpSec}. It accounts both for the contact line receding caused by capillary effects amplified by evaporation and for the motion of the bubble edge\cite{JFM21} described by using the experimental data (\autoref{fig:radii}).
From the numerical simulation, one obtains the contact line speed $U_{cl}$ and the microlayer evolution $\delta(r,t)$.
\autoref{fig:simul} shows the numerical results on the microlayer profiles at selected time moments. The microlayer appears as a thin layer of liquid trapped between the wall and the liquid-vapor interface during the bubble edge motion caused by the bubble growth. The microlayer profile consists of a growing in time ridge located near the contact line followed by a flatter and longer film showing a bumped shape.

Consider first the ridge. Its appearance is typical of dewetting phenomenon\cite{Edwards16}, i.e. retraction of a liquid film on a non-wettable wall. The dewetting ridge formation is explained as follows. A dry spot (and thus the triple contact line) appears at the microlayer. Once the microlayer starts to form, the capillary forces tend to reduce the interfacial vapor-liquid area, thus causing the contact line receding that sweeps the liquid that previously belonged to the microlayer. The net mass of liquid being evaporated in the ridge region is less than the mass of liquid being swept by the contact line motion for micrometric films\cite{JFM22}. High viscous stress in the microlayer prevents fluid flow into the film and the liquid gets accumulated near the contact line thus forming a ridge. It thus grows over time.
\begin{figure}[ht]
	\centering
	\includegraphics[width=6.5cm]{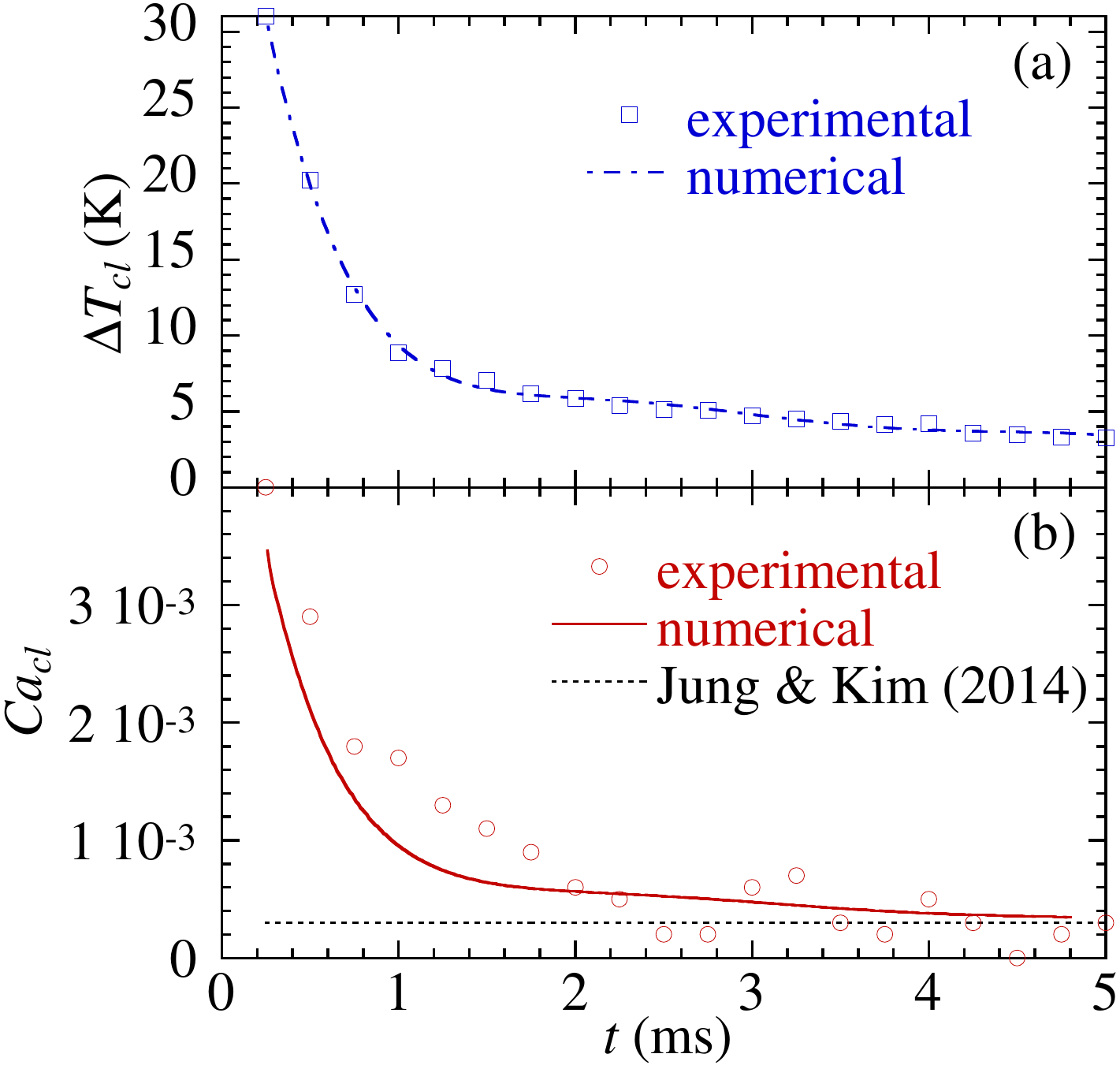}
	\caption{(a) Superheating $\Delta T_{cl}(t)=\Delta T[r=r_{cl}(t)]$ measured with the IR camera at the instantaneous position of the contact line $r_{cl}(t)$ (\autoref{fig:radii}) and its fit used for the numerical simulation. (b) Experimental (characters, obtained from \autoref{fig:radii}) and numerical (solid line) evolution of the dimensionless contact line speed $Ca_{cl}$. For comparison, we show the $Ca_{cl}$ value measured under homogeneous heating conditions\cite{Jung14}.}	\label{fig:Cacl}	
\end{figure}

The formation of a dewetting ridge is linked to the contact line motion, characterized by the capillary number $Ca_{cl}=\mu U_{cl}/\sigma$, with $U_{cl}=d r_{cl}/d t$ being the contact line speed. The theory\cite{JFM22} shows that it is mainly defined by the wall superheating at the contact line $\Delta T_{cl}=T_{w,cl}-T_{sat}$, where $T_{w,cl}$ stands for the wall temperature at the contact line. \autoref{fig:Cacl} presents the evolution of $\Delta T_{cl}$ and $Ca_{cl}$ during the contact line receding. The experimental and numerical evolutions of $Ca_{cl}$ show a good agreement. A strong correlation between $Ca_{cl}$ and $\Delta T_{cl}$ can be observed, with both featuring a decay over time. Since the mass evaporation rate at the contact line is proportional to $\Delta T_{cl}$, one can therefore understand the dewetting phenomenon as being controlled by the evaporation in the contact line vicinity.\cite{JFM22} The experimental $Ca_{cl}$ is about 10 times higher at the initial stage of contact line receding compared to its final stage where $Ca_{cl}$ value corresponds to the case of homogeneous wall heating.\cite{Jung14} The origin of such a difference is the IR laser heating mode, which promotes a highly non-uniform $\Delta T(r)$. In the beginning of evolution, the contact line radius is small so it situates at the strongly heated wall part. When the contact line recedes, it moves to the colder wall part so its speed decreases accordingly.

The dewetting ridge is observed in a large majority of bubble growth simulation studies resolving microlayer.\cite{Guion18,Urbano18,Bures21a} The essential feature of the present work is that, thanks to consideration of nanoscale effects (\autoref{ExpSec}), we are able to reproduce the contact line dynamics. The dewetting ridge was observed for slightly thicker films in another evaporation geometry.\cite{PRF16} One knows that it necessarily occurs at film dewetting, both with evaporation and without.\cite{JFM22,EPL23,Edwards16} However, it has not yet been observed experimentally in microlayer. Probably, the reason is the slope limitation of interferometry discussed in \autoref{maxSlope}. \autoref{fig:simul} shows that the slopes that occur within the ridge are high, much larger than the slope of the remaining microlayer part; the flat ridge portion is too small to be detected.

Consider now the remaining microlayer part that forms a bump. Its origin was discussed in \autoref{secbump}. The initial microlayer profile, the same as in \autoref{fig:theoretical}, is displayed in \autoref{fig:simul}a for comparison. One can see the advantage of the numerical approach over the theory: it shows the microlayer profile near the contact line and the ridge. The simulation also shows the microlayer evaporation corresponding to the thickness reduction, which is nearly uniform in agreement of the assumption advanced in \autoref{secbump}. One observes a good agreement between the experimental and simulated microlayer shapes in the vicinity of the bump, cf. \autoref{fig:simul}b. In this region, the interface slope is small enough (\autoref{maxSlope}) to be measured in our experiment.

\section{Conclusion}

Boiling is a ubiquitous phenomenon widely used in industrial applications. It is well known that during the bubble growth in boiling, the inertial effects often cause a hemispherical bubble shape with a liquid layer formed between the bubble and the heater. Because of its microscopic thickness, it is called microlayer. In this work we discuss its physical origin and apply advanced experimental methods. In particular, we use for the first time the white-light interferometry and discuss its advantages and limitations. We define the maximum interfacial slope, above which the measurements cannot be performed. This limitation equally applies to the laser interferometry.

In spite of the inertia-controlled bubble growth, the microlayer dynamics turns out to be governed by the viscous and surface tension effects. A hypothesis was emitted in the literature that the microlayer can be seen as a Landau-Levich film deposited by the bubble edge receding during the bubble growth. In this work we show that the theory based on this hypothesis agrees with the experiment. We have also explained the physics of bump in the microlayer shape observed mainly in the experiments with localized heating.

We develop a theory to explain the growth dynamics of the dry spot under the bubble (i.e. the receding of the contact line caused by the microlayer dewetting). It is an important issue in boiling, e.g. for the understanding of the CHF. We develop a theory accounting for several physical phenomena acting at nanoscale, which explains the experimentally observed contact line dynamics. This theory describes a formation of a liquid ridge along the contact line, which has not been yet observed by interferometry. For this, we provide an explanation based on the maximum observable interfacial slope defined above.

\section{Experimental and numerical approaches}\label{ExpSec}

\threesubsection{Experimental setup}\\
\begin{figure}[ht]
	\centering
	\includegraphics[width=11cm]{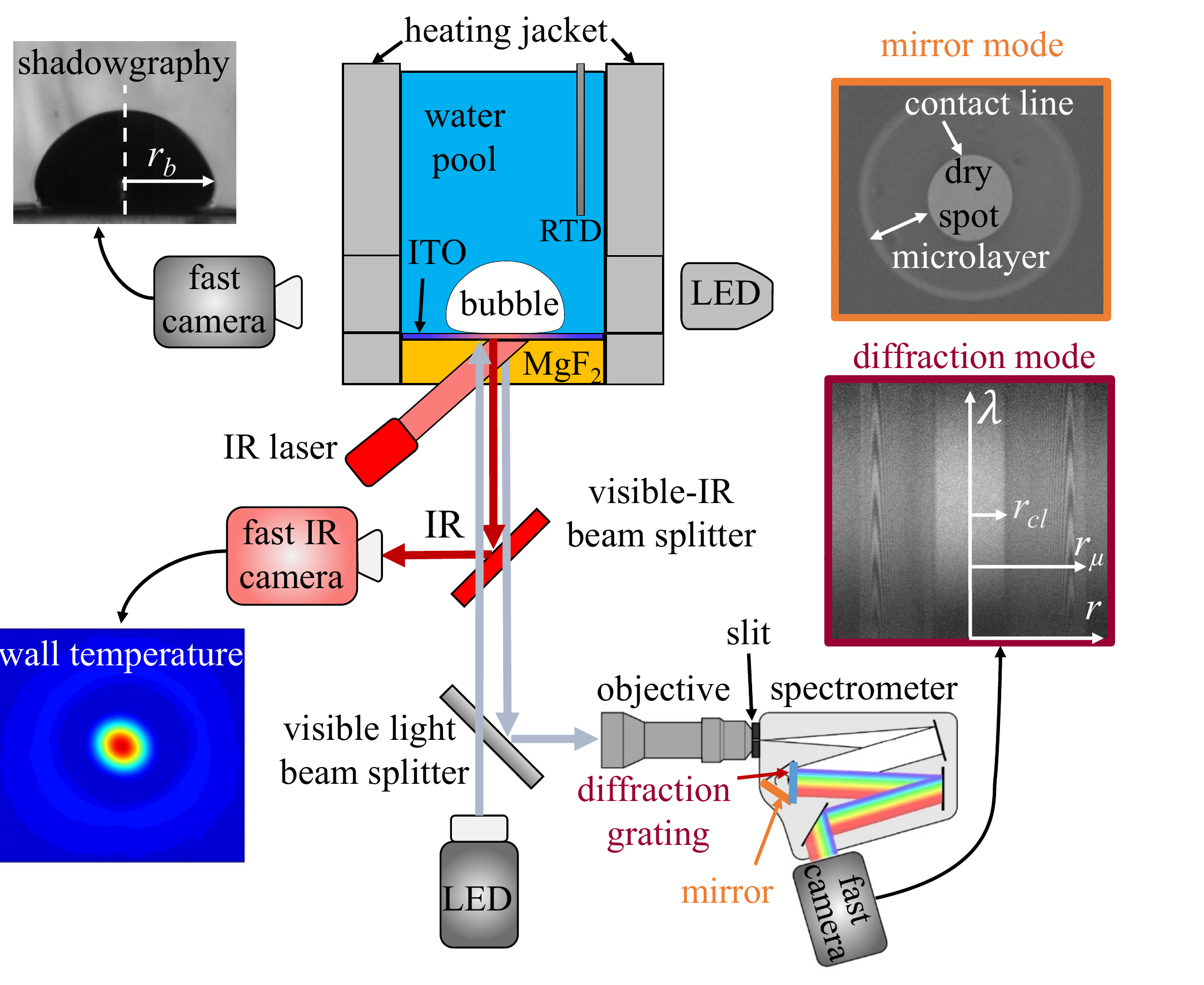}
	\caption{Schematics of the experimental installation and typical images filmed by the cameras.}\label{fig:Setup}	
\end{figure}

In our pool boiling experiment (\autoref{fig:Setup}), the boiling cell at atmospheric pressure is filled with the ultra-pure water treated with Millipore Milli-Q Integral 5 apparatus and thermally regulated at $100^\circ$C with a heating jacket connected to the thermal bath. A resistance temperature detector (RTD) is placed inside the water pool to monitor the liquid temperature; the experiment can only be started after several hours of temperature equilibration. Four lateral transparent portholes provide a means for the sideview shadowgraphy performed with Photron SA3 fast camera and telecentric collimated LED light source (Opto Engineering LT
CL HP 024-W) positioned at the opposite sides of the cell. The heater is a $\SI{950}{nm}$-thick indium-tin oxide (ITO) deposited by radio frequency magnetron sputtering on a magnesium fluoride (MgF$_2$) optical window. MgF$_2$ is transparent  both to visible and infrared (IR) light (95\% transmittance within 0.4-\SI{5}{\micro m} bandwidth) whereas ITO is transparent to visible but absorbs IR waves. The growth of a single bubble at a time is triggered by using a continuous-wave IR laser of $\SI{1.2}{\mu m}$ wavelength (Changchun New Industries Optoelectronics FC-W-1208B-10W) directed to the bottom of the MgF$_2$ window at a carefully selected nucleation site (i.e. a surface micrometric defect). Beyond the defect, the ITO surface is optically smooth with $\simeq\SI{10}{nm}$ roughness measured with AFM. The IR laser beam of $\sim\SI{1.5}{mm}$ diameter is absorbed by the ITO thus providing its localized heating.

The heart of the installation is WLI. A collimated LED white-light source (Thorlabs Solis-3C) illuminates the fluid from below. The waves reflected from liquid-vapor, liquid-ITO and MgF$_2$-ITO interfaces interfere. They are captured by the 150-\SI{600}{mm} focal length objective (Sigma F5-6.3 DG OS HSM Contemporary) installed at the entrance of a spectrometer (Horiba iHR550) with a fast high-resolution gray-scale camera (Phantom v2011) at its exit. The spectrometer can function in two modes. In the diffraction mode, a thin slit at the entrance of spectrometer defines a scanning line that (due to the \emph{in situ} adjustment) passes through the bubble center so its direction coincides with that of radial $r$ axis shown in \autoref{fig:Microlayer}. The light arriving from the slit is dispersed into wavelength by a diffraction grating inside the spectrometer, producing a fringe map captured by the camera. One obtains an intensity distribution $I_{exp}(r,\lambda)$ of the fringe pattern (right middle image in \autoref{fig:Setup}) where the ordinate axis in the figure represents the wavelength $\lambda$ and the abscissa is the coordinate $r$ (the axial symmetry of the fringes is carefully checked). The vapor has a lower index of refraction than the liquid so the dry spot appears brighter at the center of image and the dry spot radius $r_{cl}$ can be measured. The microlayer can also be distinguished thanks to the presence of interference fringes along its extent. The WLI calibration is discussed in the SI.

In the second (mirror) mode of the spectrometer, the slit is fully opened and the diffraction grating in the optical path is replaced with a motor. One obtains the 2D liquid-vapor distribution on the bubble base (right upper image in \autoref{fig:Setup}) thanks to the difference in refraction indices. Both the dry spot and microlayer can be directly observed to check their circular symmetry. To implement LI in the same setup, the LED light source can be replaced with a laser that serves also for the optics alignment.

A visible-IR beam splitter (hot mirror) is transparent to visible light but reflects the IR waves as measured by FTIR. The IR radiation emitted by the ITO is thus captured by the IR camera (FLIR X6901sc) operating in the 3-\SI{5}{\micro m} bandwidth. The camera thus images the IR radiation emitted by the ITO film (bottom left image in \autoref{fig:Setup}). By using the ITO emissivity measured by FTIR, the intensity of radiation is then related to the wall temperature through the \emph{in situ} pixel-wise calibration, see the SI.

The spatio-temporal variation of the microlayer thickness, wall temperature, and the temporal evolution of bubble characteristic sizes (\autoref{fig:radii}) are measured simultaneously and synchronously at \SI{4000}{fps} by these three methods.
\vspace*{\baselineskip}

\noindent\threesubsection{Numerical approach}

The problem is described in the 2D film cross-section of the liquid film shown in \autoref{fig:Microlayer}c. It uses the ``one-sided'' formulation, where the vapor-side hydrodynamic stress and the heat flux into the vapor at the interface are neglected compared to those on the liquid side. Therefore, the vapor pressure $p_v$ is spatially homogeneous. The solid wall is isothermal. Its temperature $T_w=T_{sat}+\Delta T$ is larger than the saturation temperature $T_{sat}=T_{sat}(p_v)$.

The generalized lubrication theory \cite{JFM22,EPL23} is used to describe the hydrodynamic flow and the microlayer thickness $\delta=\delta(r,t)$ in 2D.
\begin{figure}[ht]
	\centering
  \includegraphics[width=4cm]{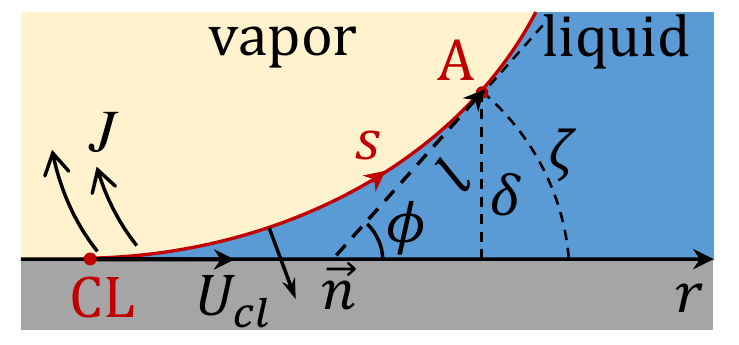}
  \caption{Sketch of the curved interface including the contact line.}\label{fig:wedge}
\end{figure}
The theory uses the parametric interface description in terms of the curvilinear coordinate $s$ that runs along the interface (\autoref{fig:wedge}), with $s=0$ at the contact line. Therefore, the following geometrical relations hold:
\begin{subequations}\label{param}
\begin{align}
{\partial \delta}/{\partial s}&=\sin \phi,\label{paramh}\\
{\partial r}/{\partial s}&=\cos \phi,\label{paramx}
\end{align}
\end{subequations}
where $\phi$ is the local interface slope. First, $\delta$ is described as a function of $s$, and then $r=r(s)$ is determined with \autoref{paramx}. The major convenience of this parametrization is the simplicity of rigorous expression for the local curvature $K$,
\begin{equation}\label{K}
K=\frac{\partial\phi}{\partial s}=\frac{1}{\cos\phi}\frac{\partial^2\delta}{\partial s^2}.
\end{equation}

At an interfacial point A, the tangent line and the solid surface form a straight wedge with the opening angle equal to the local interface slope $\phi$. The length of the intercepted arc $\zeta$ is $\phi l$, where $l$ is the radius of the straight wedge (\autoref{fig:wedge}) and is related to $\delta$ through $\delta=l\sin\phi$. Therefore, $\zeta=\delta\phi/\sin\phi$.
The interfacial pressure jump
\begin{equation}\label{eq:pJumpPr(s)}
\Delta p \equiv p_v-p_l=\sigma K -{J^2}( \rho_v^{-1} - \rho_l^{-1} ),
\end{equation}
accounts for the vapor recoil effect, where $K$ can be expressed with \autoref{K} and $J$ is the local mass flux assumed positive at evaporation.

Due to the thinness of the microlayer, heat conduction becomes the primary energy exchange mechanism, and it can be considered stationary due to its low thermal inertia. The liquid temperature is assumed to have a linear variation along the arc $\zeta$, ranging from $T_w$ to the interfacial temperature $T^i$. This assumption is based on the rigorous thermal analysis of straight wedges, where heat flow is radial \cite{Anderson1994}. The heat flux supplied to the vapor-liquid interface is spent to vaporize the liquid. As a result, the energy balance at the interface can be expressed as:
\begin{equation}\label{eq:J(s)}
  J = {k(T_w-T^i)}/({\zeta {\cal L}}).
\end{equation}
The value of $T^i$ is impacted by the Kelvin, vapor recoil and molecular-kinetic effects\cite{SWEP22}:
\begin{equation}\label{eq:TintK-pr-Ri}
T^i = T_{sat}[ 1 + {\Delta p}/({{\cal L}\rho_l})  + {J^2}( \rho_v^{-2} - \rho_l^{-2})/({2{\cal L}}) ] +  R^i J {\cal L},
\end{equation}
where
\begin{equation}
\label{eq:Ri}
    R^i = \frac{2-f}{f}\frac{T_{sat} \sqrt{2\pi R_v T_{sat}} (\rho_l-\rho_v) }{2 {\cal L}^2 \rho_l \rho_v }.
\end{equation}
is the kinetic interfacial thermal resistance with $f$, the accommodation coefficient that we assume to be unity. Here $R_v$ is the specific gas constant and $\rho_v$ is the vapor density. $T^i$ deviates from $T_{sat}$ only near the contact line, where the mass flux is large.

The governing equation
\begin{equation}\label{eq:GEA}
  \frac{\partial \delta }{\partial t}\cos\phi + \frac{\partial }{\partial s}\Bigg\{ \frac{1}{\mu G(\phi )}\Bigg[ \frac{\zeta}{2}( \zeta + 2 l_s )\frac{\partial \sigma }{\partial s}+ \frac{\zeta^2}{3}(\zeta+3l_s)\frac{\partial\Delta p}{\partial s} \Bigg] - U_{cl}\zeta \frac{F(\phi)}{G(\phi)} \Bigg\} =  - \frac{J}{\rho_l},
\end{equation}
is written in the frame of reference of the contact line receding at speed $U_{cl}$. Here, $l_s$ is the hydrodynamic slip length. Since the interface temperature $T^i$ is not constant very close to the contact line, (cf. \autoref{eq:TintK-pr-Ri}), the Marangoni effect needs to be accounted for,
\begin{equation}\label{eq:MarEff}
  {\partial \sigma }/{\partial s} =  - \gamma {\partial T^i}/{\partial s},
\end{equation}
where $\gamma=-d\sigma/d T$ is generally positive. The functions \begin{equation}
F(\phi)=\frac{2\phi^2}{3}\frac{\sin{\phi}}{\phi-\sin\phi\cos\phi}, \quad G(\phi)=\frac{\phi^3}{3}\frac{4}{\sin\phi\cos\phi-\phi\cos2\phi},
\end{equation}
are the correction factors to the conventional lubrication theory with $F(\phi\to 0)=1$ and $G(\phi\to 0)=1$. Evidently, the conventional theory reduces to \autoref{eq:energybalance} when the hydrodynamic flow (i.e. the term in curly brackets in \autoref{eq:GEA}) is neglected.

The set of governing equations (\ref{K}, \ref{eq:pJumpPr(s)}, \ref{eq:GEA}) is solved for $s\in(0, s_m)$, where $s_m$ is the end of the solution interval, i.e. the maximum $s$ value; its definition will be discussed below.

The set is of 4th order, and thus requires four boundary conditions. At the contact line, the geometry relationship implies
\begin{align}
& \delta= 0, \label{eq:BC1}\\
&{{\partial \delta}}/{\partial s}= \sin\theta_{micro}, \label{eq:BC2}
\end{align}
where $\theta_{micro}$ is the microscopic contact angle at the nanometer scale, which is regarded as the static contact angle controlling the wetting conditions. The pressure finiteness at the contact line expresses the solution regularity:
\begin{equation}\label{eq:BC3}
\left. {{\partial \Delta p}}/{\partial  s} \right|_{s \to 0}= 0.
\end{equation}

Another equation determines the contact line speed $U_{cl}$ as a part of the solution. The relation between $U_{cl}$ and the evaporation flux at contact line\cite{JFM22}
\begin{equation}\label{eq:Jcl}
J(s\to 0)=\frac{U_{cl}F(\theta_{micro})}{\dfrac{G(\theta_{micro})}{\theta_{micro}\rho_l}+\dfrac{l_s{\cal L}\theta_{micro}}{\mu k}\gamma}.
\end{equation}

At $s=s_m$, the microlayer should match the bubble interface assumed to be of constant curvature
\begin{equation}
\left. {{\partial \phi}}/{\partial  s} \right|_{s = s_m} = (\beta r_b)^{-1}, \label{eq:BC4}
\end{equation}
which is a curvature of the bubble edge that provides a transition between the hemispherical dome and the microlayer. The bubble edge has a much higher curvature than the hemispherical dome of the radius $r_b$ (the numerical variation $r_b(t)$ comes from the experiment), so $\beta\ll 1$. The value obtained in \autoref{secbump} $\beta=0.084$ is used. Similarly to the lubrication description of the liquid film deposition in the \citet{LL42} and Taylor bubble \cite{Bretherton,JFM21} problems, our formulation produces the curvature saturation to $(\beta r_b)^{-1}$ as $s\to s_m$.

One more condition is necessary for the imitation of matching of the microlayer to the bubble contour, i.e. the height $\delta_m\equiv\delta(s_m)$ at which the matching is performed. From the physical point of view, the flows near the ridge and near bubble edge are decoupled because of the absence of flow in between, so a multitude of profiles are possible for given four boundary conditions (from the mathematical point of view, because one of them is fixed at a finite $s_m$ value instead of infinity); the fifth selects one profile. The value of $\delta_m$ should be much larger than the microlayer thickness (to provide the curvature saturation) but not too high to avoid the unrealistic interfacial slopes $\geq 90^\circ$ that can appear because of the constant curvature boundary condition \eqref{eq:BC4}. We set $\delta_m(t)\simeq 0.3r_c$.

Thanks to the curvature saturation at $s\to s_m$, the solution is almost independent of the $s_m$ value. For this reason, instead of its self-consistent determination, we use the following approximate procedure. First, we mention that $r_m\equiv r(s_m)=r_b-r_{cl}$. It varies as a function of time: $\dot r_m=\dot r_b-U_{cl}$, where $\dot r_b(t)$ is obtained from the experimental data. Finally, by using the geometry constraints \eqref{param}, one can approximately convert the changes in $x_m$ and $\delta_m$ into the change in $s_m$ as
\begin{equation}
\dot s_m= \sqrt{\dot r_m^2 + \dot\delta_m^2 }.
\end{equation}

Our IR thermography measurements reveal a nearly homogeneous temperature over the microlayer extent. Therefore, we impose an uniform $\Delta T$ over the heater that changes only in time to reflect transient $\Delta T$ at the contact line as measured experimentally (\autoref{fig:Cacl}a). The temporal $\Delta T$ variation is a fit of the experimental data.

The computational algorithm is similar to that of\cite{JFM22}. The only difference is the use of remeshing at each time step, so the meshing fits the moving boundaries with the (approximate) conservation of mesh spacing while the number of nodes decreases. To find the values of $\delta(t)$ at the node positions corresponding to time moment $t+\Delta t$, the linear interpolation is used.

The constant physical parameters for water at saturation at 1 bar were taken. Two reasonable values for the parameters, the slip length $l_s=\SI{10}{nm}$ (which corresponds to the above mentioned wall roughness) and $\theta_{micro}=10^\circ$, were chosen.

\section*{Supporting Information} \par 
Supporting Information (both a text file and a video) is available from the Wiley Online Library or from the author.
\section*{Acknowledgements} \par 
We are grateful to I. Moukharski, V. Padilla and C. Rountree from SPEC/CEA Paris-Saclay for the help with the experiments and R. Abadie from SPC/CEA for supplying us with the ultra-pure water. CT would like to thank DM2S/CEA and the CFR program of CEA for the financial support of his PhD. XZ acknowledges the PhD scholarships of CNES and the CEA NUMERICS program, which has received funding from the European Union's Horizon 2020 research and innovation programme under grant agreement No 800945 - NUMERICS - H2020-MSCA-COFUND-2017.

%
\bibliographystyle{elsarticle-num-names}
\bibliography{ContactTransf,Taylor_bubbles,PHP}
\end{document}